\documentclass[12pt]{article}
\hoffset = - 0.5 truecm \voffset=-1.4 truecm
\def\beq{\begin{equation}}   \def\eeq{\end{equation}}
\def\bea{\begin{eqnarray}}  \def\eea{\end{eqnarray}} \def\nn{\nonumber}
\def\noi{\noindent} \def\beeq{\begin{eqnarray}}
\def\eeeq{\end{eqnarray}}
\def\lsim{\raise0.3ex\hbox{$<$\kern-0.75em\raise-1.1ex\hbox{$\sim$}}}
\def\gsim{\raise0.3ex\hbox{$>$\kern-0.75em\raise-1.1ex\hbox{$\sim$}}}

\usepackage{graphicx}

\usepackage{epsfig}

\renewcommand{\theequation}{\thesection.\arabic{equation}}

\newcounter{hran} \renewcommand{\thehran}{\thesection.\arabic{hran}}

\def\bmini{\setcounter{hran}{\value{equation}}

          \refstepcounter{hran}\setcounter{equation}{0}

          \renewcommand{\theequation}{\thehran\alph{equation}}\begin 
{eqnarray}}

\def\bminiG#1{\setcounter{hran}{\value{equation}}

\refstepcounter{hran}\setcounter{equation}{-1}

\renewcommand{\theequation}{\thehran\alph{equation}}

\refstepcounter{equation}\label{#1}\begin{eqnarray}}

\def\emini{\end{eqnarray}\relax\setcounter{equation}{\value{hran}} 
\renewcommand{\theequation}{\thesection.\arabic{equation}}}

\begin{document}

\begin{flushright}

            \today\\

	   LPT-Orsay 10-57\\

\end{flushright}

\vspace{1.cm}

\begin{center}

\vbox to 1 truecm {}

{\large \bf NLO Calculation of Prompt Photon}

\par \vskip 3 truemm

{\large \bf  Production in DIS at HERA}

\vskip 2 truecm

{\bf P. Aurenche$^1$, Rahul Basu$^2$, M. Fontannaz$^3$} \vskip 3 truemm

{\it $^1$ LAPTH, Universit\'e de Savoie, CNRS, \\ 
BP 110, Chemin de Bellevue, 74941 Annecy-le-Vieux
Cedex, France}

\vskip 3 truemm
{\it $^2$ The Institute of Mathematical Sciences, Chennai 600 113,  
India}

\vskip 3 truemm

{\it $^3$ Laboratoire de Physique Th\'eorique, UMR 8627 du CNRS,\\
Universit\'e Paris XI, B\^atiment 210, 91405 Orsay Cedex, France}
\vskip 2 truecm

\begin{abstract}
We present a NLO calculation of prompt photon production in DIS. The calculation
involves direct, fragmentation and resolved contributions. It is performed in
the virtual-photon proton center-of-mass system. A comparison of the theoretical
results with HERA data is carried out. 
\end{abstract}

\end{center}

\newpage
\pagestyle{plain}

\baselineskip=22 pt

\section{Introduction}
\hspace*{\parindent}
The prompt photon production in DIS $eP$ collisions is an interesting  
reaction involving, in the theoretical description provided by QCD,  
several perturbative and non perturbative quantities related to virtual and  
real photons. At first sight the subprocess associated with the  
prompt photon production appears particularly simple. It is the  
Compton effect of a virtual photon on a quark with a real photon and  
a quark in the final state: $\gamma^* + q \to \gamma + q$. The  
theoretical description of this process only requires the knowledge  
of the quark distributions in the proton and the calculation of Higher  
Order (HO) QCD corrections, the latter opening the way to a  
quantitative comparison with experimental data.\par

Actually, as always with photons, the situation is quite complex. We  
also have to consider the reaction in which the initial virtual  
photon fluctuates into a state made of collinear quarks and gluons  
described by the virtual photon structure function. In the final  
state also, a large-$p_\bot$ quark can radiate a real collinear  
photon, a process which involves a perturbative part --the  
bremsstrahlung of the photon-- and a non perturbative part, and which  
is described by the photon fragmentation function. Therefore the  
complete description of the prompt photon production requires the  
calculation of four classes of processes. Two classes involving the virtual  
photon structure function with the final photon either directly  
coupled to a quark of the hard subprocess (the resolved-direct  
process) or produced through the final fragmentation of a parton (the  
resolved-fragmented process). The two other classes involve the  
direct coupling of the virtual photon to a quark of the hard  
subprocess with a direct final photon (the direct-direct process) or  
with a photon fragment of a parton (the direct-fragmented process).  
All four processes corresponding to these four topologies having, as  
we shall see below, the same order of magnitude. Photons can also be  
emitted by the lepton line. We do not consider this contribution in  
this paper. It can be obtained, from ref. \cite{0r,4new}.\par

On the experimental side, H1 and ZEUS \cite{1r,2r} have measured the  
prompt photon inclusive cross sections for which no QCD HO  
calculations exist. H1 also measured the production of a prompt  
photon associated with a jet. For this latter reaction a HO  
calculation does exist \cite{4new} which concerns the direct  
production. The fragmentation process is also taken into account and  
many observables are discussed in ref. \cite{4new}. It is the aim of  
this paper to present QCD HO calculations for the four processes of  
inclusive production of prompt photons in DIS $eP$ collisions, and to  
compare the theoretical predictions with H1 and ZEUS data. We do not  
consider the production of a photon and a jet. But the approach of  
this paper is directly applicable to this latter case.\par

An important point in the definition of the prompt photon cross  
section is that of the reference frame in which the large-$p_\bot$  
photon is observed. There are two standards at HERA, the photon-proton 
(center of mass) CM system (hadronic system) and the laboratory frame. In the  
first one the observed large-$p_\bot$ particle is produced, at lowest  
order, via $2 \to 2$ subprocesses. This frame has been used in almost  
all large-$p_\bot$ reactions at HERA. We shall work in such a frame  
in the present paper. It is natural in photoproduction in which the  
almost real initial photon is collinear with the initial electron.  
For instance the photoproduction of prompt photons, which is the $Q^2  
\to 0$ limit of the DIS reaction, has been studied in this frame \cite 
{4r,5r,6r}. A lower cut-off in $p_\bot$ is necessary for perturbative QCD 
to be valid. 
The forward production of a $\pi^0$, or a jet, in the DIS  
reaction has also been studied with this low-pt cut-off which picks $2 \to 2$  
subprocesses (and their HO corrections) to produce large-$p_\bot$  
forward partons with the aim to test the importance of the BFKL   dynamics
\cite{7r,8r,9r,10r,10kkm,11r,12r,13r}. A detailed comparison with experimental
results was performed and a good agreement of theoretical results with the data
was found \cite{10r,10kkm,13r}. \par

The situation is different in the laboratory in which a $2 \to 2$  
subprocess is no more necessary to produce a large-$p_\bot$ particle.  
The transverse momentum of the observed photon may come from that of  
the virtual photon $q_\bot = \sqrt{Q^2(1- y)}$ ($y$ being the  
inelasticity) through the basic DIS subprocess $\gamma^* +$ quark $\to 
$ quark, followed by the emission of the photon from the final quark.  
The authors of references \cite{14r} have proposed to use this frame  
as a way to attain the quark into photon fragmentation function, and  
have done a detailed study of this reaction at order ${\cal O} 
(\alpha^3)$. \par

Therefore the choice of a frame amounts to emphasize a given  
subprocess and the related non perturbative quantities. To be short  
one could say that the laboratory frame emphasizes the fragmentation  
part, whereas the $\gamma^*$-$P$ frame emphasizes the virtual photon  
structure function and the fragmentation part. Unfortunately only  
isolated photons have been observed at HERA, corresponding to a  
strong suppression of contributions involving fragmentation  
functions. \par

Of course it is not necessary to fix the transverse momentum and the  
rapidity of the observed photon in the $\gamma^*$-$P$ frame. This can  
be done in the laboratory. We only have to check that the subprocess  
in the $\gamma^*$-$P$ frame involves a large $p_\bot$ scale. For  
instance H1, when observing forward (in the laboratory) and large-$p_ 
\bot$ (in the laboratory) $\pi^0$, requires a cut on the transverse  
momentum of the $\pi^0$ in the hadronic frame \cite{7r}. We shall  
follow this procedure in this paper. \par

In the next section we give details on the HO calculations in the  
hadronic frame. In section 3 we present results for the inclusive  
cross section as a function of $E_\bot^\gamma$, $y^\gamma$ and $Q^2$.  
We discuss the effect of isolation (isolated prompt photons are  
measured by H1 and ZEUS). Then we examine, in section 4, the  
possibility to relate our $\gamma^*$-$P$ results with the H1 and ZEUS  
results obtained with no $p_\bot$ cut in the hadronic frame. We will  
identify laboratory phase space regions in which this cut is not  
necessary to insure a large-$p_\bot$ in the $\gamma^*$-$P$ frame.  
Section 5 is a conclusion.

\section{Technical details}
\hspace*{\parindent}
The programs containing a fragmentation function, which describe the  
resolved-fragmented and the direct-fragmented processes, can be  
immediately obtained from the programs of $\pi^0$ production in DIS  
$eP$ reactions \cite{11r,13r}. The only change is that of the  
fragmentation functions; we now use the Bourhis-Fontannaz-Guillet  
(set II) \cite{15r} fragmentation functions. \par

The programs describing the direct-direct and resolved-direct  
processes can be obtained from the preceeding ones by selecting the  
subprocesses with a final gluon and changing the charges and colour  
factors. Thus we obtain the cross sections corresponding to the  
emission of a photon.\par

In all these programs the HO corrections associated with a $2\to 3$  
subprocesses are calculated in dividing the phase space of the third  
final parton, which can be soft, in three parts: a small cylinder  
around the initial momenta (in the hadronic frame in which the  
virtual photon and the proton one collinear), a small cone around the  
two final hard partons, and the rest of the phase space. The cylinder  
allows us to treat the collinear and soft singularities associated  
with the initial partons. The cones allow us to calculate the  
collinear and soft singularities associated with the final partons.  
In the rest of the phase space the $2 \to 3$ cross sections have no  
singularities and the integration is performed by a Monte Carlo  
method  \cite{16r}. \par

This approach is described in details in references \cite{17r}. It is  
at the root of all the programs of the PHOX-FAMILY \cite{18r}. What  
is peculiar to DIS $eP$ reactions is the presence of a virtual photon  
structure functions. Several problems are raised by these functions ;  
for instance the implementation of the $\overline{\rm MS}$  
factorization scheme, their NLO evolution, their parametrizations. A  
detailed discussion of all these points has been given in reference  
\cite{12r}. Here one can only keep in mind that the available NLO  
parametrization are given in the $\overline{\rm MS}$ factorization  
scheme. \par

As only isolated photons are observed in the HERA experiments, we  
have implemented an isolation criteria. We will use a cone criterion  
in the hadronic frame requiring no more than $\varepsilon E_\bot^\gamma$  
hadronic transverse energy in a  cone of radius $R^\gamma$  
surrounding the photon, with $\varepsilon = .1$ and $R^\gamma= 1$ (a  
value $\varepsilon = .111$ will be used when comparing theoretical  
results with H1 and ZEUS data\footnote{The experimental cuts require that the
photon carries at least 90\% of the total energy of the jet of which it forms a part
corresponding to $\varepsilon = .111$  with our conventions.}).\par

The photon transverse energy can be fixed in the hadronic frame, in  
which we require a minimum value of the latter for a perturbative  
approach to be valid. This cut also eliminates contributions from the  
collinear subprocess $q + \gamma^* \to q$ followed by the  
bremsstrahlung of a final photon. The photon kinematics can also be  
fixed in the laboratory, as it is done by H1 and ZEUS. In this case,  
after a boost to the hadronic frame, the photon transverse energy may  
be very small. We suppress the contributions corresponding to this  
configuration by requiring again a minimum transverse energy in the  
hadronic frame. In the next section we present results obtained ``in  
the laboratory frame'' in order to be close to the kinematics used by  
the HERA experiments, but with a cut in the hadronic frame.\par

The calculations of the photon-jet cross sections could proceed in a  
similar way. We just have to introduce a jet algorithm in the  
routines calculating the $2 \to 3$ subprocesses, specifying how two  
final partons are combined to form a jet. The interest of the photon- 
jet cross sections is due to the simultaneous measurement of the  
photon $p_\gamma$ and jet $p_J$ four momenta, which allows us to  
define the invariant mass $m_{\gamma J}^2 = 2 p_\gamma \cdot p_J$.  
Whatever the frame in which these momenta are measured, perturbative  
QCD calculations are valid if $m_{\gamma J}^2$ is large enough and we  
do not need any more a $p_{\bot\gamma}$ cut in the hadronic frame  
\cite{4new}.

\section{Cross sections in the hadronic frame}
\hspace*{\parindent}
In this section we present the cross sections corresponding to the  
four topologies described in the introduction, and we consider non  
isolated and isolated photons.  \par

We adopt kinematical parameters close to those of the HERA  
experiments to make later comparisons with H1 and ZEUS easier. The  
beam energies of the proton and lepton are respectively 920~GeV and  
27.6~Gev leading to $\sqrt{S_{ep}} = 318.7$~GeV. The inelasticity $y = {q 
\cdot p \over \ell \cdot p}$ is taken in the range $.1 < y < .7$ and  
the photon virtuality $Q^2$ in the range 4 GeV$^2 < Q^2 < 100$~GeV$^2 
$. ($P$, $\ell$ and $q$ are the four momenta of the proton, the  
initial lepton and the virtual photon). The cross section involving the  
exchange of the $Z$-boson are neglected.\par

The rapidity of the photon in the laboratory frame is $1.8 > y_\gamma > - 1.2 
$ and the transverse energy 10 GeV $> E_{\bot\gamma} > 3$~GeV. In  
this section we do not consider the other cuts put by H1 and ZEUS on  
the scattering angle of the lepton, on the momentum of the outgoing  
lepton and on the invariant mass $(P + q - p_\gamma)^2 = W_x^2$ that  
we shall introduce in section 4 when comparing with data. Finally, in  
order to stay in a perturbative regime, we require the photon to  
have, in the hadronic frame, a minimum transverse momentum $p_{\bot 
\gamma}^* > 2.5$~GeV. \par

We use the CTEQ6M distribution functions \cite{21r}, the parton  
distributions in the virtual photon of ref. \cite{12r} and the BFG  
photon fragmentation functions (set II) \cite{15r}. We work with $N_f  
= 4$ flavors. \par

\begin{figure}[htb]
\vspace{9pt}
\centering
\includegraphics[width=3.5in,height=2.5in]{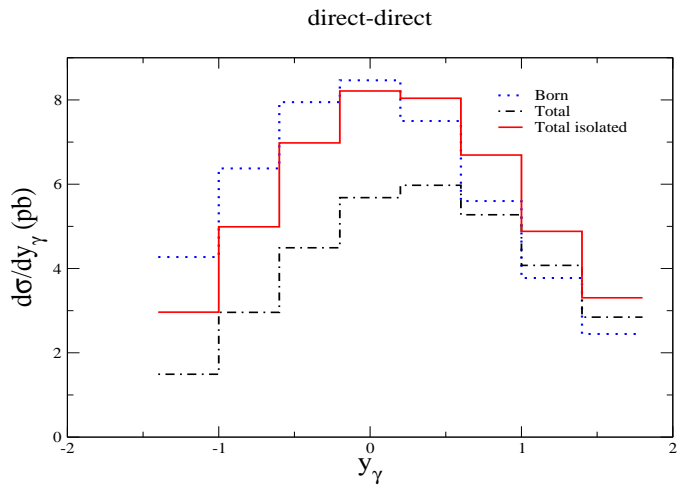}
\caption{The direct-direct cross section $d\sigma/dy_\gamma$}
\label{fig:1}
\end{figure}

Fig. 1 to Fig. 4 present the cross sections $d\sigma /dy_\gamma$  
corresponding to the four topologies discussed in the introduction.  
We must keep in mind that these results are factorization scheme  
dependent. They depend on the initial factorization, final  
factorization and renormalization scales. Only the sum of the four  
partial cross sections has a physical meaning.  
The scales used in this section are $M = C \sqrt{Q^2 + (p_{\bot  
\gamma}^*)^2}$ for the proton factorization scale, $M_\gamma =  \sqrt 
{Q^2 + (C_\gamma p_{\bot\gamma}^*)^2}$ for the virtual photon  
factorization scale, $M_F = C_F \sqrt{Q^2 + (p_{\bot\gamma}^*)^2}$  
for the fragmentation factorization scale and $\mu = C_\mu \sqrt{Q^2  
+ (p_{\bot\gamma}^*)^2}$ for the renormalization scale, with $C = C_ 
\gamma = C_F = C_\mu = 1$. All our calculations are performed in the $ 
\overline{\rm MS}$ factorization scheme. In particular the HO terms  
with a direct final photon are obtained from one-loop calculations  
from which final collinear singularities are subtracted with the $ 
\overline{\rm MS}$ convention.

\begin{figure}[htb]
\vspace{9pt}
\centering
\includegraphics[width=3.5in,height=2.5in]{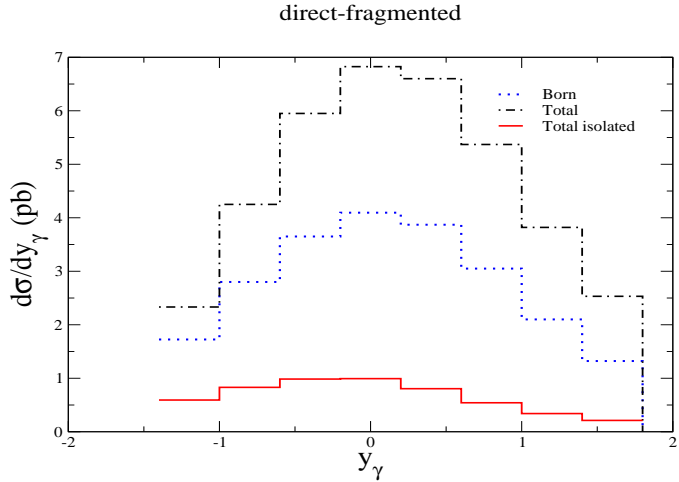}
\caption{The direct-fragmented cross section $d\sigma/dy_\gamma$}
\label{fig:2}
\end{figure}

\begin{figure}[htb]
\vspace{9pt}
\centering
\includegraphics[width=3.5in,height=2.5in]{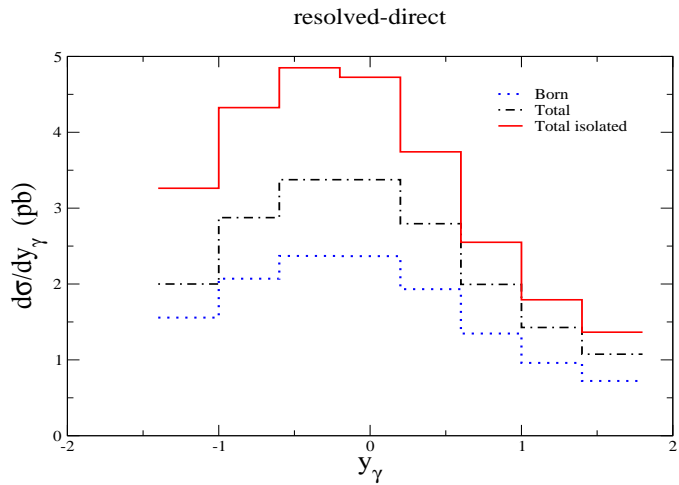}
\caption{The resolved-direct cross section $d\sigma/dy_\gamma$}
\label{fig:3}
\end{figure}

\begin{figure}[htb]
\vspace{9pt}
\centering
\includegraphics[width=3.5in,height=2.5in]{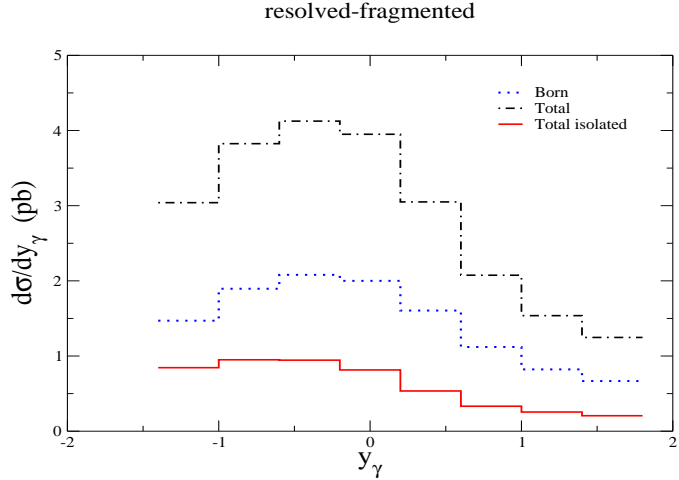}
\caption{The resolved-fragmented cross section $d\sigma/dy_\gamma$}
\label{fig:4}
\end{figure}

Let us first consider the production of inclusive non isolated photons and let  
us start with the Born terms (Figs.~1-4). As expected the largest  
contribution comes from the direct-direct reaction. However the other  
reactions are not negligible. The direct-fragmented contribution is a  
factor 2 below the direct-direct one, and each of the resolved contributions  
another factor two below so that the direct-direct contribution represents only
half of the full cross section.\par

The HO corrections are important. We display in the figures the Total = Born +
HO cross section for each class. The HO corrections almost double the Born  
terms for the components involving fragmentation. This is due to the  
large number of subprocesses participating in the cross sections. For  
the contributions with a final direct photon, the pattern is  
different. For instance in the direct-direct case the HO corrections  
are very negative for $y_\gamma$ smaller than $.5$. This produces an  
important correction of the cross section especially in the negative  
$y_\gamma$ range. The peculiarities of the direct-direct reaction is  
the presence of only one subprocess at the Born level, namely $ 
\gamma^* + q \to \gamma + q$, and the presence of strong kinematical  
constraints (parton distributions in the photon and fragmentation  
functions replaced by delta-functions).\par

When the isolation is switched on, we obtain an interesting pattern  
of Born terms and HO corrections. As expected we have a strong  
decrease of the Born sections involving fragmentation functions of  
about a factor six, or even larger in the forward direction.  
Therefore these Born cross sections are almost negligible compared to  
the direct Born cross section on which the isolation has no effect.  
For the HO order corrections the decreases are less pronounced  
compared to the Born terms and they even increase in the cases of  
final direct photons; for instance by a factor 2 for the resolved-direct  
case. This increase of the isolated total cross section, for the case where the
photon is directly produced in the final state (Figs.~1 and 3), has already  
been observed and discussed in ref. \cite{22r} which studies the  
prompt photon production in hadron-hadron collisions. It is due to  
the fact that some HO final collinear contributions are subtracted from  
the HO direct cross section to build the fragmentation functions. The  
remaining collinear terms are negative and the subtracted HO direct  
contribution increases when these negative terms are cut by the  
isolation. Of course this effect is factorization scheme dependent  
and demonstrates once more that only the sum of the cross sections  
has a physical meaning (the total cross section must decrease when  
the isolation is switched on). \par\vskip 5 truemm

\begin{figure}[htb]
\vspace{9pt}
\centering
\includegraphics[width=3.5in,height=2.5in]{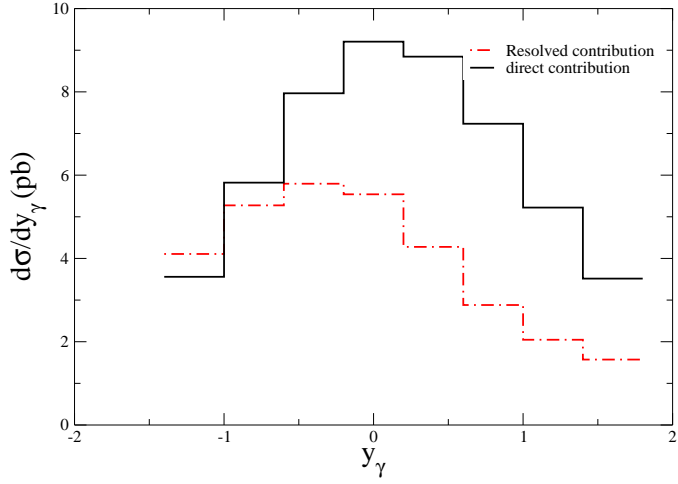}
\caption{Resolved and direct isolated cross sections $d\sigma/dy_ 
\gamma$.}
\label{fig.5}
\end{figure}

\begin{figure}[htb]
\vspace{9pt}
\centering
\includegraphics[width=3.5in,height=2.5in]{fig6.eps}
\caption{Resolved and direct isolated cross sections $d\sigma/dE_{\bot 
\gamma}$.}
\label{fig.6}
\end{figure}

In conclusion, we observe  
that the resolved-direct isolated cross section is larger than the  
non isolated one (Fig.~3). This is also true for the direct-direct cross  
section with a very different behavior in rapidity (Fig.~1). The isolated  
resolved-fragmented and the direct-fragmented cross sections are small  
compared to the isolated resolved-direct and direct-direct  
contributions, but not totally negligible.\par

In Figs.~5 and 6 we finally show the contributions of the resolved  
and the direct processes for the cross section $d\sigma /dy_\gamma 
$ and $d\sigma /dE_{\bot \gamma}$ (isolated case). These figures  
illustrate the non negligible contributions of the resolved components
specially for negative values of the photon rapidity.

\section{Comparison with experimental results}
\hspace*{\parindent}

Unlike what was done until now for the production of large $p_{\bot}$
hadrons or jets, the  H1 and ZEUS
experiments do not put, in the case of prompt photons, kinematical cuts in the  
hadronic system and it is easy to check that their laboratory cuts do  
not protect the hadronic system $p^*_{\bot\gamma}$ against small  
numerical values. Thus the possibility of performing a perturbative  
calculation is not ensured. The values of $Q^2$ and $y$ being fixed  
by the observation of the outgoing lepton, we have to calculate the  
expression
\beq
\label{1e}
\int dQ^2 dy \int {dx \over x} G(x) {\ell^{\mu\nu} t_{\mu\nu} \over  
Q^4} \delta^{(4)} \left ( p + q - p_4 - p_3\right ) d\vec{p}_{\bot 4}  
\ dy_4 \ d\vec{p}_{\bot 3} \ dy_3 \ ,
\eeq

\noindent the definition of the momenta being given in Fig.~7a  
corresponding to the laboratory frame.\par

The delta-function can be written in terms of the light cone  
variables, transverse momenta and rapidities.
$$2 \delta^{(2)} \left ( \vec{q}_\bot - \vec{p}_{4\bot} - \vec{p}_{3 
\bot}\right ) \delta \left ( x P^{(+)} - {Q^2 \over \ell^{(-)}} - p_ 
{\bot 4} \ e^{y_4} - p_{\bot 3}\ e^{y_3}\right ) \delta \left ( y  
\ell^{(-)} - p_{\bot 4} \ e^{-y_4} - p_{\bot 3}\ e^{-y_3}\right )$$
\bea
\label{2e}
&&= {2 \over P^{(+)} \ell^{(-)}} \delta^{(2)} \left ( \vec{q}_\bot -  
\vec{p}_{4 \bot} - \vec{p}_{3\bot}\right ) \delta \left ( x - {Q^2 y  
\over 2 P\cdot q} - {p_{\bot4} e^{y_4} + P_{\bot 3} e^{y_3} \over p^ 
{(+)}}\right ) \nn \\
&&\delta \left ( y - {p_{\bot 4} e^{-y_4} + p_{\bot 3} e^{-y_3} \over  
\ell^{(-)}}\right )
\eea

\noi with
\begin{eqnarray*}
&&P^{(+)} = P_0 + P_z = 2 P_z \\
&&\ell^{(-)} = \ell_0 - \ell_z = 2E
\end{eqnarray*}

\noi and $p^{(+)} = x P^{(+)}$.

 From (\ref{2e}) we obtain the following constraints
\bea
\label{3e}
&&e^{y_4} = {p_{\bot 4} \over y \ell^{(-)} - p_{\bot 3}\ e^{-y_3}}  
\qquad \hbox{and} \nn \\
&&\vec{p}_{\bot 4} = \vec{q}_\bot - \vec{p}_{\bot 3}\ .
\eea

We see that we can ensure a minimum value of $\widehat{s} = {\cal O}  
(p_{\bot 3} p_{\bot 4})$, the subprocess center-of-mass energy  
squared, with the requirement $\sqrt{Q^2(1-y)} = |\vec{q}_\bot| < | 
\vec{p}_{\bot 4}|$. \par

To discuss the constraints on $y_4$ let us consider the extreme case  
where $\vec{p}_3$ is parallel to $\vec{p}_4$ ($\widehat{s} = 0$).  
With the definition $q_\bot = p_{\bot 3}+ p_{\bot 4}$ and with $y_3 =  
y_4$ we get from the constraints (\ref{2e})
\bea
\label{4.3e}
&&x = x_{B_j} = {Q^2 \over y S_{ep}} \nn \\
&&e^{y_4} = {\sqrt{Q^2(1- y)} \over 2y\ell^0} \ .
\eea

\noi The first result is the standard constraint associated with the  
$q + \gamma^* \to q$ subprocess; the second result shows that $y_4$  
cannot be large if $Q^2$ is small and $y$ bounded from below. This  
can be rewritten as $q_\bot = e^{y_4} 2 \ell^0 y$. With the H1 cut $y_ 
{min} = .05$ we obtain $p_{\bot min} = 2.76\ e^{y_4}$, which shows  
that the collinear configuration ($\widehat{s} = 0$) does not  
contribute to the cross section in the small $p_\bot$ and large $y_4$  
domain.

\begin{figure}[htb]
\vspace{9pt}
\centering
\includegraphics[width=3.5in,height=1.5in]{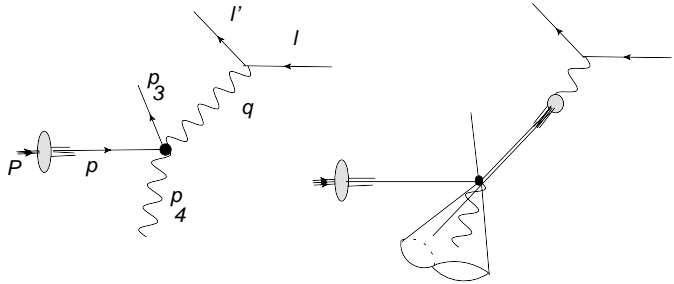}
\caption{a) direct-direct kinematics in the LAB frame. b) idem for  
the resolved-direct subprocess.}
\label{fig.7}
\end{figure}

\subsection{Comparison with H1 data}
\hspace*{\parindent}
To determine the exact phase space domain in which our calculation is  
valid, we explore, for large value of $y_4$ or for small values of  
$Q^2$, the sensitivity of the cross section to $p_{\bot\gamma}^*$- 
cuts. In this study we switch to the exact H1 cuts which include the  
conditions $E'_e > 10$~GeV, $153^{\circ} < \theta_e < 177^{\circ}$ and  
$W_x > 50$~GeV. Now the values of the inelasticity $y$ are bounded  
below by $.05 < y$, and we have $4 < Q^2 < 150$~GeV$^2$. We also implement  
the isolation criterion in the laboratory frame by requiring a  
limited hadronic energy $E_\bot^h$ in a cone of radius $R^\gamma$ around the  
photon. This isolation is different from that defined in the hadronic  
frame since parton 3 (fig. 7) can now be inside the isolation  
cone because of the transverse boost provided by the virtual photon. We use the
isolation parameters $\varepsilon = .111$ and $R^  \gamma = 1$. We perform our
exploration of the ``safe'' domains by   studying the direct-direct cross
section. A typical result that we   obtain for the small $Q^2$ domain $4 \leq
Q^2\  \lsim\ 10$~Gev$^2$ is given in table 1. All other H1 parameters and cuts
being as given above.

\begin{table}[htb]
\begin{center}
\begin{tabular}{|c|c|c|}
\hline
&& \\
\hbox to 1 truecm {}${\bf p}_{\bot\gamma}^{\bf *}-${\bf cut}\hbox to  
1 truecm {} &\hbox to 1 truecm {}{\bf Born}\hbox to 1 truecm {} & 
\hbox to 1 truecm {}{\bf NLO}\hbox to 1 truecm {} \\
&& \\
\hline
&& \\
2.5\  {\rm GeV}&6.60 &7.14 $\pm$ .05 \\
1.5 \ {\rm GeV}&7.50 &7.64 $\pm$ .05 \\
.5 \ {\rm GeV}&7.60 &7.62 $\pm$ .05 \\
&& \\
\hline
\end{tabular}
\end{center}

\caption{Integrated direct-direct cross section in the range $4 \leq  
Q^2 \leq 10$~GeV$^2$ in picobarns.}
\label{tab1}
\end{table}

We notice, as expected, a very small dependence on $p^*_{\bot\gamma}$- 
cut. Exploration of the range $10 \leq Q^2 \leq 20$~GeV$^2$ leads to a much  
pronounced dependence on $p_{\bot\gamma}^*$-cut, the cross section  
(Born term) varying from $4.82$ to $7.2$~pb. Interesting results are  
also given by H1 for the range $4 \leq Q^2 \leq 40$~GeV$^2$, the  
laboratory transverse momentum $q_\bot = \sqrt{Q^2(1-y)} \leq $ being  
not too large compared to $E_{\bot\gamma}$. We explore for this $Q^2$ 
range the large $y_\gamma$ range $1.4 \leq y_\gamma \leq 1.8$ and  
find for the cross section a smaller variation then the one shown  
in table 1.\par

To conclude we find small variations of the cross section with $p^*_ 
{\bot\gamma}$-cut in the small $Q^2$-range or in the large $y_\gamma$  
range. This incites us to compare our predictions with H1 data in the  
above domains. The H1 collaboration provides us with data in the  
domain $A$ : $4 < Q^2 < 10$~{\rm Gev}$^2$, $3 < E_{\bot \gamma} < 10 
$~GeV, $-1.2 < y_\gamma < 1.8$, in the domain $B$ : $4 < Q^2 < 40$~GeV$^2 
$, $6 < E_{\bot \gamma} < 10$~GeV, $-1.2 < y_\gamma < 1.8$, and in  
the domain $4 < Q^2 < 40$~GeV$^2$, $3 < E_{\bot\gamma} < 10$~GeV,  
$1.4 < y_\gamma < 1.8$. In these three domains the variations with $p_ 
{\bot\gamma}^*$-cut of the cross section is small, especially domains  
$B$ and $C$. In the first domain the stability of the resolved-direct  
($rd$) cross section is not as good as than that of the direct-direct ($dd 
$) contribution of table 1. It varies for 4.45 to 5.91 for $p_{\bot 
\gamma}^*$-cut ranging from 2.5 GeV to .5 GeV.\par

Our predictions are compared with data in table 3 and details of the  
contributions are given in table 2. In these three domains we set $p^*_{\bot\gamma}$-cut
= 1.0 GeV.
\begin{table}[htb]
\begin{center}
\begin{tabular}{|c|cc|cc|cc|}
\hline
{\bf Domain} &\multicolumn{2}{|c|}{\bf dd contribution } & 
\multicolumn{2}{|c|}{\bf rd contribution }&\multicolumn{2}{|c|} 
{\bf df contribution }\\
&{\bf Born} &{\bf NLO} &{\bf Born} &{\bf NLO} &{\bf Born} &{\bf NLO}\\
\hline
&&&& \\
A &1.26 &1.27 &.59 &1.18 &.0785 &.192  pb/GeV$^2$\\
&&&&&& \\
B &.74 &.67 &.32 &.56 &.0475 &.098  pb/GeV\\
&&&&&& \\
C &3.72 &4.66 &1.02 &1.88 &.153 &.370  pb\\
&&&&&& \\
\hline
\end{tabular}
\end{center}
\caption{Details of the $dd$, $rd$ and $df$ contributions.}
\label{tab2}
\end{table}

\vskip 3 truemm

\begin{table}[htb]
\begin{center}
\begin{tabular}{|c|c|c|c|}
\hline
{\bf Domain} &{\bf Data } &{\bf f}$_{\bf had}$ &{\bf dd + rd + df  
NLO }\\
\hline
&&& \\
A &2.48 $\pm$ .21 $\displaystyle{+ .34 \atop{-.41}}$ &.87 &2.30  pb/GeV$^2$ \\
&&& \\
B &1.78 $\pm$ .25 $\displaystyle{+ .46 \atop{-.60}}$ &.97 &1.29  pb/GeV \\
&&& \\
C &4.38 $\pm$ 1.26 $\displaystyle{+ 1.04 \atop{-1.75}}$ &.77 &5.32  pb\\
&&& \\
\hline
\end{tabular}
\end{center}
\caption{Comparison between data and theory.}
\label{tab3}
\end{table}

\noi In the three domains\footnote{The data in domains B and C have been  
obtained by subtraction between the 4 $<$ Q$^2 <$ 150 and 40 $<$ Q$^2  
\leq$ 150. Statistical errors are added in quadrature and we keep the  
largest systematic errors.} we find a good agreement between theory and  
experiment. Note that the last column of table 3 has been obtained by  
multiplying the predictions of table 2 by the hadronic correction  
factors $f_{had}$ given by H1 \cite{1r}. The small 
resolved-fragmented contribution has not been taken into account;  we  
estimate this contribution to be approximatively equal to 6~\% of the  
total cross sections. Note also that the contributions due to the  
emission of photons by the lepton line is not \par






\noi taken into account. The H1 collaboration has estimated this  
contribution \cite{1r}. It is almost negligible ($\sim 5\ \%$) in  
domain $A$, it represents about $13\ \%$ of the theoretical cross  
section (table~3) in domain $B$, and is totally negligible in domain  
$C$. \par

\subsection{Comparison with ZEUS data}
\hspace*{\parindent}

The ZEUS collaboration measured the isolated photon cross section at the energy
$\sqrt{S_{ep}} = 318$~GeV using the following cuts: 
$E'_e > 10$~GeV, $139.8^{\circ} < \theta_e < 171.9^{\circ}$ and  
$W_x > 5$~GeV. The kinematical domain covered by the experiment is:
$10 < Q^2 < 350$~GeV$^2$, $4 < E_{\bot \gamma} < 15 $~GeV and  
$-.7 < y_\gamma < .9$. The data are quoted in various bins for each of these
three variables with the other two variables integrated over the whole indicated
range. From our previous dicussion on H1 data it appeared that stability of our
predictions under the $p_{\bot \gamma}^*$ cut-off could be achieved for  low
$Q^2 < 10$~{\rm Gev}$^2$, or large photon rapidity $y_\gamma  > 1.4 $ with
moderate $Q^2$ values, however none of these domains can be extracted from the
published ZEUS data. Restricting to large $E_{\bot \gamma}$ values, $e.g.$ 
$E_{\bot \gamma} > 10$~GeV, we have tested that it is imposssible to obtain
stability in the full $Q^2$ range of ZEUS. A relative stability of our
predictions is achieved, however, when one restricts the $Q^2$ range between 10 and
20~GeV$^2$ integrating over the whole range for the other variables: we find a
cross section (with only the d-d and d-f contributions) decreasing from .164
down to .152~pb/GeV$^2$, when decreasing the cut-off from 2.5 to 1. GeV (the d-d
HO are not stable). Multiplying the result by a factor 2 to roughly take into
account the resolved contribution, we obtain a cross section of .37 pb/GeV$^2$
(including a contribution of .045 pb/GeV$^2$ from the lepton line) compatible
with the value of .414 $\pm$ .035 (stat.)~pb/GeV$^2$ given by ZEUS.

\section{Conclusions}

A complete calculation of the isolated photon cross section in deep inelastic
scattering has been presented at the next to leading logarithmic order in QCD.
Isolation can be imposed in the hadronic center of mass frame or  in the
laboratory frame. The calculation includes four classes of processes depending
on whether the photon is coupled directly to the hard process or through
structure or fragmentation functions and it is valid if the momentum of the
photon in the hadronic center of mass frame is large enough to prevent the
2$\rightarrow$1 process, $q \rightarrow q + \gamma$, to occur in a collinear
configuration. Unfortunately the H1 and ZEUS collaborations for their isolated
prompt photon studies do not impose a transverse momentum cut-off in the
hadronic center of masss frame but in the laboratory frame. This is unlike what
was done  for their studies on particle or jet production.  The comparison
between our calculation and the data can be performed only in a very restricted
domain of the data: small $Q^2$ and/or large photon rapidity.  It would be
interesting if data could be available with a transverse momentum cut in the
hadronic frame: a comparison with large transverse momentum $\pi^0$ production,
will then be possible and it is an exercise always worth making as is done in
hadronic colliders. The basic mechanisms of photon and $\pi^0$ production are
different and so are the higher order corrections: a detailed comparison of
theory with data in the case of photon production with many kinematical
variables at hand ($Q^2$, $E_{\bot \gamma}$, $y_\gamma$, $x_{\rm Bj}$) will be
very constraining, as it was for $\pi^0$ production.


\begin{thebibliography}{99}
\bibitem{0r} RAPGAP, H. Jung, Comput. Phys. Commun. {\bf 86} (1995) 147.
\bibitem{4new} A. Gehrmann-De Ridder, G. Kramer and H. Spiesberger,  
Nucl. Phys. {\bf B578} (2000) 326.
\bibitem{1r} H1 collaboration, F.D. Aaron et al., Eur. Phys. J. {\bf  
C54} (2008) 371.
\bibitem{2r} ZEUS collaboration, S. Chekanov et al., Phys. Lett. {\bf  
B687} (2010) 16.
\bibitem{4r} ZEUS collaboration, S. Chekanov et al., Eur. Phys. J.  
{\bf C49} (2007) 511.
\bibitem{5r} H1 collaboration, F.D. Aaron et al., Eur. Phys. J. {\bf  
C66} (2010) 17.
\bibitem{6r} M. Fontannaz, J. Ph. Guillet and G. Heinrich, Eur. Phys.  
J. {\bf C21} (2001) 303.\\
M. Fontannaz and G. Heinrich, Eur. Phys. J. {\bf C34} (2004) 191.
\bibitem{7r} H1 collaboration, A. Aktas et al., Eur. Phys. J. {\bf  
C36} (2004) 191.\\
H1 collaboration, A. Aktas et al., Eur. Phys. J. {\bf C46} (2006) 27.
\bibitem{8r} ZEUS collaboration, S. Chekanov et al., Eur. Phys. J.  
{\bf C52} (2007) 515.
\bibitem{9r} G. Kramer and B. P\"otter, Phys. Lett. {\bf B453} (1999)  
295.
\bibitem{10r} A. Daleo, C.A. Garcia Canal and R. Sassot, Eur. Phys.  
J. {\bf C33} (2004) 404.\\
A. Daleo, D. de Florian and R. Sassot, Phys. Rev. {\bf D71} (2005) 034013.
\bibitem{10kkm}
B.A. Kniehl, G. Kramer and M. Maniatis, Nucl. Phys. {\bf B711} (2005)  
345, E : {\bf B720} (2005) 231.
\bibitem{11r} P. Aurenche, R. Basu, M. Fontannaz and R.M. Godbole,  
Eur. Phys. J. {\bf C34} (2004) 277.
\bibitem{12r} M. Fontannaz, Eur. Phys. J. {\bf C38} (2004) 297.
\bibitem{13r} P. Aurenche, R. Basu, M. Fontannaz and R.M. Godbole,  
Eur. Phys. J. {\bf C42} (2005) 43.
\bibitem{14r} A. Gehrmann-De Ridder, T. Gehrmann and E. Poulsen, Eur.  
Phys. J. {\bf C47} (2006) 395.
\bibitem{15r} L. Bourhis, M. Fontannaz, J. Ph. Guillet, Eur. Phys. J.  
{\bf C2} (1998) 529.
\bibitem{16r} S. Kawabata, BASES integration package, Comp.Phys.Comm. 
88(1995) 309
\bibitem{17r} T. Binoth, J. Ph. Guillet, E. Pilon, M. Werlen, Eur.  
Phys. J. {\bf C16} (2000) 311.\\
P. Aurenche, L. Bourhis, M. Fontannaz, J. Ph. Guillet, Eur. Phys. J.  
{\bf C17} (2000) 413.
\bibitem{18r} P.~Aurenche, T.~Binoth, M.~Fontannaz, J.~Ph~Guillet, G.~Heinrich,
E.~Pilon, M.~Werlen, http://lappweb.in2p3.fr/PHOX-FAMILY/main.html.
\bibitem{19r} S.D. Ellis and D.E. Soper, Phys. Rev. {\bf D48} (1993)  
3160.
\bibitem{20r} G. Salam, Towards Jetography, arXiv:0906.1833 [hep-ph].
\bibitem{21r} J. Pumplin, D.R. Stump, J. Huston, H.L. Lai, P.  
Nadolsky and W.K. Tung, JHEP {\bf 0207} (2002) 012.
\bibitem{22r} S. Catani, M. Fontannaz, J. Ph. Guillet, E. Pilon, JHEP
 0205 (2002) 028
\end{thebibliography}
\end{document}